\documentclass[aps,prl,twocolumn,showpacs,superscriptaddress,groupedaddress]{revtex4}
\usepackage{graphicx}
\usepackage{dcolumn}
\usepackage{bm}
\usepackage{amssymb}
\begin{document}
%\preprint{APS}
\preprint{APS/MA-FR-JB}

\title{MANYBODY ASPECTS OF GRAVITY IN COMPACT STARS}

\author{Mofazzal Azam}
\email{mofazzal.azam@gmail.com}
\affiliation{Centre for theoretical physics, Jamia millia Islamia, New Delhi}

\author{Jitesh R. Bhatt}
\email{jiteshbhatt.prl@gmail.com}
\affiliation{Theory Division, Physical Research Laboratory, Ahmedabad}

\author{M. Sami}
\email {samijamia@gmail.com }
\affiliation {Centre for Theoretical Physics, Jamia Millia Islamia, New Delhi}
%\email {jiteshbhatt.prl@gmail.com}
\date{\today}

\begin{abstract}

Compact stars such  as neutron stars and black holes are gravitationally bound many body systems. We investigate
%make some critical comments on
the importance of short and long range part of gravity for such systems. From our analysis, we conclude that the true essence of gravity
lies with the long range nature of the interaction. At the end we show how these arguments in the manybody theory consistently leads to
Dvali-Gomez picture of a black holes as a collective bound state of long wavelength gravitons.

%We make some critical comments on the importance of short range and long range part of the interaction for a gravitionally 
%bound manybody system.

\end{abstract}


\maketitle

\setlength{\abovedisplayskip}{9pt}

\setlength{\belowdisplayskip}{9pt}

\newpage
The distinguishing feature of gravity is that it is universal as well as long range. Short range interaction implies interaction of a particle with other particles only in its neibourhood and thus the number of particles it  interacts with is proportional to local density.
%Coulomb interaction is long range but charge neutrality of manybody system renders it short range. 
The long range of gravity in manybody context implies that every particle in the system interacts with every other particle and,  thus, there is interaction at all length scales including the
system size scale. We will  be concerned here with manybody systems,  stellar like  compact objects, such as neutron stars and black holes, bound by gravity, and consider the role of short and long range 
part of the interaction in the stability of such systems using classical as well as quantum theory. We will see that in the manybody context, it is the long range
part of the interaction that plays the dominant role. It is not hard to understand why it is so. Long range involves large number of particles within the range of the interaction and thus,  even if
the two-particle interaction energy is low, for a large system they add up to a very large number and easily dominate over the short range contributions.
We consider here only the Newtonian part and no mention will be made of general relativity until at the end. Except for black holes, qualitative features of all gravitationally bound manybody systems can be understood using Newtonian part of gravity. In the case of neutron stars, general relativity introduces some change in the 
mass and radius but the qualitative features are unchanged. The neutron-neutron repulsive potential also plays a role in fixing the mass and radius of the neutron stars. The discussion given in this paper can easily be carried over to include this contribution and thus will not be considered. Our level of discussion is qualitative and sometimes even heuristic. We are more concerned about the
physics of the problem and thus, quite often we ignore "order one" quantities. In tune with modern usage of terminology in high energy physics, we frequently use
the word "infra-red (IR)", for long range and "ultra-violet (UV)" for short range. They refer to the wavelengths of quanta corresponding to the energetics of the problem.\\
\\
 We begin the discussion with stability analysis of a simple quantum mechanical problem.
 %Create a new first level heading
We encounter two types of instabilities in the Quantum Mechanical treatment of gravitationally bound systems - one is of UV origin and the other is of IR origin, and they are in some sense independent of each other. For the sake of illustration,  let us consider a gravitationally bound system of two point particles of mass $ M$ and $ m$. We assume that  $ M>> m $. It is a problem similar to that of a high $Z$ hydrogenic atom  but with the difference that the
 interaction is of gravitational origin.
% Let us also suppose that they are too close and thus the kinetic energy is relativistic.
% Assume $ M>> m$. 
Thus,  for the low lying orbits kinetic energy of the lighter particle is relativistic. We can write,
\begin {eqnarray}
{\mathcal{E}} \approx c|p| -G\frac {M m}{R}. 
%\nonumber
\end {eqnarray}
Using the uncertainty relation to  
replace $ p $ by $ p\approx \hbar/R $ and  we obtain,
%$ p\approx \frac {\hbar}{R} $, and thus get,
\begin {eqnarray}
{\mathcal{E}}=\frac {c\hbar}{R}-G\frac {M m}{R }=\frac {m_{p}^{2}- M m} {G R},
%\nonumber\\
%where ~~~m_{p}^{2}=\frac {c\hbar}{G}\nonumber
\end {eqnarray}
where, $ m_{p}^2=c\hbar /G$.
%\frac {c\hbar}{G} $.
When $\sqrt{Mm}> m_p $,  the limit  $ R~\rightarrow~0$,  leads to   ${\mathcal {E}}~\rightarrow~-\infty $.
Therefore, the system is unstable as the ground state
is not bounded from below.
 This instability persists even if we change the Newtonian potential ($1/R $)
%$\frac{1}{R} $ 
with the Yukawa type potential $ exp (-R/\lambda) /R $.
This is because the  instability in the present case is due to the short range part of the potential energy and it persists even when we remove the long range part of the potential by using the Yukawa type interaction. This short range or ultra-violet (UV) instability can exist even in  a  manybody system, but it can become important only at  the Planck density.

 As we explain later in the text, the long range or
the infra-red (IR) effects of gravity become dominant much before we reach the Planck density\cite{thir}.
%It is not hard to see that at planck density these UV contributions dominate the overall energy functional. The binding energy per  particle is unbounded from below and decreases to $-\infty $, as we let density $\rho $ increase towards $\infty $ . \\
\\
The UV instability that we have discussed above is also known for high $Z$ hydrogenic atoms. In this case the stability requirement puts a bound on $ Z\alpha $. For manybody Coulomb systems with overall charge neutrality and densities such that the constituent particles are relativistic,  we require bound on $ Z\alpha $ as well as $\alpha $ for the system to be stable \cite{lieb1,lieb2}.  In this case,  by stability, we mean that the binding energy per particle in the system is finite and is independent of the number of constituent particles. This kind of UV instability is also known in nuclear physics where the nucleons interact via short Yukawa potential. The stability of nuclei is assured by the existence of additional shorter range hard core repulsive potential \cite{cohe,shal}.
It should be noted that, the UV instability of gravity is important in a physical situation involving scattering of two particles having energy of the order of Planck mass. The instability in the  manybody case makes it transparent that the UV instability is connected with the local density and  not the system size. This is the lesson from the  IR turncated Yukawa type short range gravitational interaction. \\
\\
 There are some subtleties in the case of manybody
system with a gravitational interaction which we will
discuss below. First of all there is no screening of the long range gravitational potential. Unlike in the case of Coulomb/Nuclear interaction, the overall size of the system plays a major role in fixing the local densities in a gravitationally bound system.  We will argue that the UV instability in the manybody situation discussed above, in fact, is irrelevant in all physical situations. It is the IR instability which plays major role in fixing the local as well global properties of gravitationally bound manybody systems.  However, for a situation like scattering of two particles with energies of the order of Planck mass, the UV aspects of gravitational interaction still remains relevant and important. We  will first discuss the classical problem of Jeans length because , in our view, it is this length scale that is at the heart of IR behaviour of a gravitionally bound system. We consider here systems with constant density profile and investigate how this profile changes as we change the parameters of the problem. This is a bit artificial as gravitationally bounded systems always give rise to a peak in the central density profile and this feature is also generated by the infrared behaviour of gravity. However, for the theme to be discussed here this feature introduces unnecessary complications without any new insight.\\
\\
Let us first discuss a problem which is very simple but important for the underlying theme of this paper.
% simple problem from elementary text book in high school physics.
We consider a solid ball of radius $ R $, density $\rho $ and mass $ M $. We drill a surface to surface hole through the centre of the sphere and put a point particle of mass $ m $ in this thin tunnel.  The point particle at any point at distance  $ r $ from the centre is under the influence of the gravitational potential of inner sphere of radius $ r $, and the potential of the spherical shell of inner radius $ r $ and outer radius $ R $. The potential energy $\Phi (r) $ of the interaction is,
\begin{eqnarray}
\Phi (r)=\frac {2\pi}{3}G m \rho r^2-2\pi Gm\rho  R^2
% \nonumber
\end{eqnarray}
and thus the equation of motion of the point particle is,
\begin{eqnarray}
m\frac {d^2r}{dt^2}=-\frac{d\Phi}{dr}=-\frac{4\pi}{3} G\rho m r
%\nonumber
\end {eqnarray}
Thus it is simply the equation of motion of a harmonic oscillator,
\begin {eqnarray}
\frac {d^2r}{dt^2}+\frac {4\pi}{3} G\rho r=0
%\nonumber
\end{eqnarray}
with frequency $\omega=(\frac {4\pi}{3})^{1/2} \times \sqrt {G\rho} $ and time period $ T=\frac {2\pi}{\omega}=\sqrt {3\pi}\times\frac {1}{\sqrt {G\rho}} $. From now onwards quite often we will drop order one quantities and simply write $\omega=\sqrt{G\rho} $ and $ T=1/\sqrt{G\rho}$.  
%Note also that the density here is not the same as the density introduced earlier. It is more of a global quantity and should be thought of às density functional which is taken to be constant through out the material body. In astrophysics, we need to define matter density of various types reflecting various aspects of the cosmic body. First of all we have the local density $\rho (r) $ as a function of a point $ r$ within the system,  there is  the central density $\rho_c $ defined as density strictly at the central region of the spherical body, and then there is the density $\bar{\rho} (r)$, defined as the ratio of mass  $ M(r) $ contained within sphere of  radius $ r$ and the volume $ V (r)=\frac {4\pi}{3} r^3$. Our global density is more like $\bar {\rho} $. Clearly, in the case of constant density profile, all the three types of densities coincide. In fact, quite often we will simply use the word " density" without further explanation,  and it should be understood from context when it is a local quantity and when it is a global quantity. Thus $\omega $ is the inner characteristic frequency of the gravitating system. We will come back to this quantity time and again in the text.\\
\\
In the example above  we considered the gravitational field of solid sphere of constant density and found a characteristic frequency. But we did not consider the sphere to be bound gravitationally. Thus the mass density in this problem is arbitrary.  In a gravitationally bound system the density is fixed by over all size of the system and so, the problem is a bit artificial. However, it should be noted that even such an artificial system captures one important aspect of gravitational interaction: there is a lowest  frequency of oscillation in the system , and corresponding to that there is a characteristic time scale.  The essence of this time scale is understood when we consider gravitationally bound systems. Gravitationally bound manybody system has a  characteristic length scale, called Jeans' length \cite{jean,mich,clar}. We shall argue that the Jeans length is actually dictated by the IR behaviour of gravity and is connected with the frequency and time scale just introduced.\\
\\
Let us consider a spherically symmetric gas cloud of mass $ M $, of constant density $\rho $ and at pressure $ P $. The volume of gas $ V=\frac {4\pi}{3} R^3$
The total energy of the gas,
\begin {eqnarray}
E=\frac {3}{2} PV-\frac {3}{5}~\frac{GM^2}{R}.
%\nonumber
\end {eqnarray}
We want to find the thermodynamic configuration for which the spherically symmetric gas cloud would become unstable under gravity. This is analysed by finding the extremum of energy per unit volume with respect to variations in density but keeping the total mass of the gas cloud fixed. We introduce the following notations: $\epsilon(\rho)=E/V $, $ R=R (\rho) $ , $c_{s}^2=\frac {dP}{d\rho} $, $ c _s $ is the velocity of sound in the medium.
Thus the conditions for the extremum of energy density are 
\begin {eqnarray}
\frac {d\epsilon(\rho)}{d\rho}=0~~; ~~ \frac{dM(\rho)}{d\rho}=0.
%\nonumber
\end {eqnarray}
It leads to the following equation,
\begin {eqnarray}
\frac {d\epsilon(\rho)}{d\rho}=\frac {3}{2} \frac {dP}{d\rho}-\frac {16\pi}{15} G\rho R^2=0.
%\nonumber
\end {eqnarray}
Thus we have an extremum for,
\begin {eqnarray}
R=R_J\equiv \sqrt{\frac {45}{32\pi}}~ \frac {c_s}{\sqrt{G\rho}} ; ~~ c_s=\sqrt{\frac{dP}{d\rho}}.
%\nonumber
\end {eqnarray}
To investigate the nature of the extremum, we calculate the second derivative of the energy density at, $~R=R_J $,
\begin{eqnarray}
\frac {d^2 \epsilon (\rho)}{d\rho^2}=3 c_s (\frac {dc_s}{d\rho}~-~\frac {1}{6}~\frac{c_s}{\rho}).
%\nonumber
\end {eqnarray}
This is negative when the speed of sound is independent of mass density. Thus the extremum is  a maximum and fluid sphere is unstable
 for radius $ R\geq R_J$.
%= c_{s}/\sqrt {G\rho}$.
The more general condition for instability is obtained from,
\begin {eqnarray}
\frac {dc_s}{d\rho}-\frac {1}{6}~\frac {c_s}{\rho}<0.
%\nonumber
\end {eqnarray}
This gives $ c_s\sim\rho^{\eta} $, where, $o\leq\eta <1/6$. Thus from $ dP/d\rho=c_{s}^2~$, we get $ P\sim\rho^{(2\eta+1)}$. The borderline for
stability is given by $\eta=1/6$.

%This gives $c_s\sim\rho^{1/6 } $. From $ dP/d\rho=c_{s}^2~$ , we get $ P\sim \rho^{4/3} ~$. 
Thus for radius $ R\geq R_J $, and 
the equation state of the fluid, $ P\sim\rho^{(\frac{4}{3}-\epsilon)}$,   $0 <\epsilon\leq 1/3 $, the fluid sphere is gravitationally unstable. 
Note that,
\begin {eqnarray}
R_J\simeq c_s/\sqrt {G\rho}~~ \rightarrow~ R_J\simeq \frac {c_s}{\sqrt {\frac {3GM}{4\pi R_{J}^3}}}~~\rightarrow ~R_J\simeq \frac {GM}{c_{s}^2}.\nonumber\\
\end {eqnarray}
For a relativistic gas, velocity of sound is of the order of velocity light($c$) and thus in the present case
%the radius of the black hole precursor can be obtained 
simply by replacing $ c_s\rightarrow c $, we obtain $ R\simeq \frac {GM}{c^2} $. This may suggests that we should be able to express the black hole radius in the form of Jeans radius,
\begin {eqnarray}
R_g=\frac {2GM}{c^2}~\rightarrow~R_g=\frac {8\pi G \rho R_{g}^3}{3 c^2}~\rightarrow~R_g\simeq \frac {c}{\sqrt {G\rho}}. \nonumber\\
\end {eqnarray}
However, the concept of mass density within a black hole is problematic or rather vague and therefore, this radius should be understood as the radius of some stellar configuration just before black hole formation.Thus, there seem to exist a pattern in these unstable configurations leading us to view classical black holes as extreme limit of such configurations.

The Jeans instability arises solely due to long range (IR)  behaviour of the Gravitational interaction and is independent of the short  range (UV) behaviour.  We provide here a very simple qualitative argument. Let us make the replacement, 

\begin {eqnarray}
 \Phi_{N} (r)\equiv  -\frac {Gm_1  m_2}{r} \rightarrow\Phi(r)\equiv \Phi_{N}(r)-\Phi_{UV}(r),\nonumber
\end{eqnarray}
where,
\begin{eqnarray}
\Phi_{UV}(r)= -\frac{Gm_1 m_2}{r} \Theta (r_0 -r).\nonumber
\end{eqnarray}
The potential $\Phi (r) $, concides with the Newtonian potential for $ r\leq r_0$, but it equals to zero for $ r> r_0$. Here $ r_0$, is assumed to 
be "not too large" . This simply  means that it is much much  smaller than the system size. Latter, in the text, we will provide a rough estimate of its value.

Thus,

\begin{eqnarray}
\Phi(r)=\Phi_{N}(r)  \equiv  -\frac{Gm_1 m_2}{r} ~~~for~~r> r_0,\nonumber
\end{eqnarray}
\begin{eqnarray}
\Phi (r)=0 ~~for ~~r\leq r_0. \nonumber
%\nonumber
%=\phi_{N}(r)+\phi_{Y}(r)\nonumber
\end {eqnarray}

%It is easy to see that, for $ r \rightarrow 0$, the potential energy of two particle interaction,  $\phi_{\lambda}(r)\rightarrow -\frac {G m_1 m_2}{\lambda} $.
% and  the force  between the particles,  $ F\equiv  -\frac {d\phi_{\lambda}(r)}{dr}\rightarrow 0$. 
%The  repulsive short range Yukawa type potential, $\phi_{Y}(r) $ , modifies the Newtonian potential at short distance scale, $ r <\lambda $, but at large distance scales  $ r>>\lambda$, the potential $\phi _{\lambda}(r) $ coincides with Newtonian potential. Thus the potential $\phi_{\lambda} (r)$ is  UV finite. 
The potential $\Phi (r) $ implies that each particle interacts only with particles outside a sphere of radius $ r $ around it.
We will often refer to it as UV finite Newtonian potential. \\
Let us calculate the potential energy of the many body system due to the
short range part $ \Phi_{UV}(r) $.
% the repulsive Yukawa type interactions.
 The range of the interaction is $r_0$  and thus each particle interacts only with $N' \equiv \frac{4\pi }{3} r_{0}^3 n  \approx r_{0}^3 n $ particles.
(unlike in the case of Newtonian interaction where each particle interacts with all the $ N$ paticles in the system). 
Here the number density of particles, $ n=N/V\simeq N/R^3$,  mass density,  $\rho=nm$, mass $ M=Nm $. The interaction energy of single particle is
given by, $ -\frac{Gm^2 n r_{0}^3}{r_0}=-Gn m^2  r_{0}^2$. Thus the energy of interaction for all the $ N $ particles in the system due to the potential $\Phi_{UV}(r) $ is,
\begin{eqnarray}
E_{UV}\approx - G N nm^2 r_{0}^2.
\end{eqnarray}
%average interparticle distance, $ a\simeq n^{-1/3}$. 

%The potential energy of two particle interaction at average interparticle distance $a $ is $ (Gm^2/a)~e^{-a/\lambda} \simeq Gm^2/a\simeq Gm^2 n^{1/3}$.Each particle interacts with $\lambda^3 n $ particles in the system, and thus the repulsive potential energy per particle is, 
%\begin {eqnarray}
%\Phi_{Y}=Gm^2\lambda n^{4/3}
%\nonumber
%\end {eqnarray}
The total potential energy due to Newtonian interaction is given by,
\begin{eqnarray}
 E_N=-\frac {3}{5 }\frac {GM^2}{R}\simeq -Gm^2 N^2/R\simeq -Gm^2n^{1/3} N^{5 /3}, \nonumber\\
\end{eqnarray}
\noindent
where, we have used,  $ n^{1/3}\simeq N^{1/3}/R$.
% Dividing this expression of potential energy by the number particle $ N$, we obtain per particle potential  energy due to this potential,
%\begin {eqnarray}
%\Phi _{N}\simeq -Gm^2n^{1/3} N^{2/3}
%\nonumber
%\end {eqnarray}
The total energy of interaction due to the UV finite interaction , $\Phi =\Phi_N - \Phi_{UV} $,  is given by,
\begin{eqnarray}
E=E_N - E_{UV}\simeq -Gm^2 n^{1/3} N^{5/3} + Gm^2 N n r_{0}^2. \nonumber\\
\end{eqnarray}
Thus, the per particle interaction energy is,
\begin{eqnarray}
\epsilon =\epsilon_{N}-\epsilon_{UV} \simeq  -Gm^2n^{1/3} N^{2/3} +Gm^2n r_{0}^2.\nonumber\\
\end{eqnarray}

Note that  the potential energy per particle,  $\epsilon_{N} $, depends on the total number of particles in the system. On the other hand, $\epsilon_{UV} $, depends on the local density and is too small and negligible.
% It is relevant only at Planck scale.
 Thus  the potential energy per particle  for UV finite potential
$\Phi(r) $ is the same as that of Newtonian , $\epsilon=\epsilon_{N}-\epsilon_{UV}\simeq\epsilon_{N} $. Thus for the formation of Gravitationally bound structures UV part of the Newtonian potential is irrelevant.\\
 %because the UV finite part of the potential which we constructed by adding repulsive short range Yukawa type potential  leads to the same results. 
%The immediate precursors of supermassive or stellar mass black holes are formed at densities which are much much smaller than Planck scale and thus the UV part of Newtonian does not play any role. 
It is relevant only when there is Planck scale density fluctuations .  \\
%In the case of UV modified Newtonian gravity we have introduced cancellation of this contribution by adding a Yukawa type repulsive interation and even then we have Jeans instability,  star formation, Chandrasekhar limit of compact stars and also the black hole precursors. This is because they are all related to infrared instability of gravity in a manybody system.
It should be emphasized here that we are not proposing the UV modified gravity as an alternative to Newtonian gravity\cite{adel,salu,azam,azam1}
 but merely stressing the irrelevance of the short range part and the associated UV instability in the manybody context. 
 We will show that the UV cutoff can  be taken to be
 several $ cms $ (a range in which Newtonian gravity is well tested). Thus the  emergence of stellar structures is solely  
due to the infrared instability of gravity.
It is also not difficult to infer, in the case of compact stars, that infrared instability  leads to the emergence of "Planck scale" 
quantity,  such as the "Planck mass",  but here it should be understood more like a critical value of a parameter which defines boundary between  a stable and an unstable configuration. In $ c=\hbar=1$ unit, the Planck mass, $ m_p=1/\sqrt{G} $, is simply the inverse of the square root of gravitational coupling constant and it sets up the size of largest stable structure.

 The energy per particle of a relativistic neutron star in the case of UV finite potential $ \Phi(r) $,  is,
\begin {eqnarray}
\epsilon_{NS} \simeq c\hbar n^{1/3} - Gm^2 n^{1/3} N^{2/3} + Gm^2 n r_{0}^2.
%\nonumber
\end {eqnarray}
In this equation, the first term is the Fermi energy of a neutron in a relativistic  neutron star, the second term comes from the 
%repulsive Yukawa potential that cancels the UV attractive part of
 Newtonian potential and the third term comes from the potential $\Phi_{UV} $, that cancels the UV part of Newtonian potentials.
 %The last term is the usual Newtonian potential energy . 
The neutron star is unstable for
 $\epsilon_{NS} <0$   \cite{shap,chan1}. Thus for the stability ($\epsilon_{NS}>$0), we gets the following condition:
\begin {eqnarray}
N^{2/3}>  \frac{c\hbar}{Gm^2}+n^{2/3} r_{0}^2 =\frac {m_{p}^2}{m^2}+n^{2/3} r_{0}^2.
%\nonumber
\end {eqnarray}
%Here $ m_p= \sqrt{ \frac{c\hbar}{G} }$ is the planck mass. 
%Newtonian  gravity has been tested up to a distance scale of a micron, and thus we take $\lambda=10^{-6} cm$. 
For a typical neutron star of mass density $10^{14} ~gms/cc$, number density of neutrons, $ n\simeq  10^{38} ~neutrons/cc$,  neutron mass, $ m=1 GeV$,  Planck mass, $ m_p=10^{19} GeV $,  the first term in the equation above is $10^{38} $. For UV cutoff parameter, $ r_{0}=10~ cms $,
the second term is of order of $10^{28} $ which is ten order of magnitude less than the first term and, therefore, its contribution is negligible. The stability of neutron star is insensitive to the presence of the second term, even for the cutoff scale $ r_{0}=10~ cms $.
% The second term which is related to UV part of interaction is too small compared to the first term and as expected does not play any role in the stability of a neutron star. 
Thus the stability of a neutron star is completely determined  by the long range part of the gravitational interaction. Note the appearance of Planck mass in the formula even in the UV finite gravity. This shows that IR instability of gravity is also implicitly connected with Planck mass but there is no Planck length with the usual meaning in this case because  now we have the UV cut off.

Lesson from the above discussion is that in the case of Newtonian gravity like in Coulomb and nuclear systems, there exists UV  instability. In nuclear systems the hard core repulsive potential controls this instability. In relativistic Coulomb system, this instability leads to a bound on $ Z\alpha $ and $\alpha$ which are realistic.  In the case of gravity UV instability occurs only at densities of Planck scale. In a manybody situation such a density fluctuation is not possible unless a large macroscopic scale sub-system/system collapses to a point. However, IR instability occur at realistic densities and in some cases it leads to the formation of black hole much before  the Planck scale density is reached.

 For a gravitationally bound stellar object as emphasised earlier, it is the long range part of interaction that plays the major role in its binding. This means that it is the interaction at approximately the system-size length scale
 that determines the field configuration within the star. In a Quantum theory, 
this should correspond to virtual gravitons of system-size wavelength. As mass content within the size of a system increases, the occupation number of such gravitons should also increase. In Newtonian gravity, there is no self-interaction of the gravitational fields.
%, and thus in this case the virtual gravitons do not interact with each other.
 However, in general relativity, there is a self-interaction and thus  the virtual gravitons interact with each other. These are long wavelength low energy soft gravitons, and the interaction strength between two of them is weak. However,  their number is very large and each one of them interact with all others. Thus an individual graviton is in the collective gravitational potential of all other gravitons. This makes the coupling strong. The question is : Can this collective interaction of Gravitons in black hole lead to the formation of coherent state of the gravitons? 
%In other words, do the virtual gravitons in black hole precursors condense into bose liquid? 
Dvali and Gomez in a series of publications have taken this view and tried to explain various aspects of black holes \cite{dval1,dval2}. They introduce a quantity called graviton occupation number,

\begin{eqnarray}
N_g = \frac{cMR}{\hbar}.
%\nonumber
\end{eqnarray}

For a fixed $ R $, this quantity is maximum when the mass, $ M $ is  given by $ R= 2GM/c^2$, i.e., equal to the black hole mass for the radius $ R $. Before we discuss the black hole, let us consider a critical neutron star of mass $ M_n $ and number of neutrons $ N_n $, where for the critical neutron 
 star, $ N_n\approx  (\frac {c \hbar}{G m_{n}^2})^{3/2} $ and $ R_n \approx \frac {\hbar}{m_n c} N_{n}^{1/3} $. Here $ M_n=N_n m_n $. The (zero point) kinetic energy of the neutrons in the star is, $ \frac {c\hbar N_{n}^{4/3}}{R_n} $. Now for such a star, $ N_g = \frac{c M_n R_n}{\hbar}= N_{n}^{4/3} $. Here we have used the fact that $ R_n= \frac {\hbar}{m_n c} N_{n}^{1/3} $. Thus the total kinetic energy of system size gravitons, $ N_g\times \frac {c\hbar}{R_n}$ is of the same order of magnitude as the total kinetic energy of neutrons in the star (numerical factor of order one not considered). Let us define inertia of system size graviton
as $ m_g=\frac {\hbar}{R_n c} $ , $ p_g=\hbar/R $, being the momentum of such gravitons. Substituting, the values fot $ N_g $ and $ m_g$,
it is not hard to see that the total potential energy of self intetaction of gravitons, $-\frac {G N_{g}^2 m_{g}^2}{R_n}$ is of the same order of magnitude as that of the neutrons in the star.\\
To appreciate why we are doing this exercise, let us consider a neutron star which is increasing its mass by acreating and processing matter from its sarroundings.When its mass $ M\geq M_{crit}$, and number of neutrons in the star $ N\geq N_{crit} $, its potential energy is of the order of "$-GM^{2} /R=
-GN^{2} m_{n}^{2}/R~ $".
%$-\frac {GM^2}{R} =-\frac { G N^2 m_{n}^2}{R} $. 
Had this energy been radiated away, it will take away  the inertia and thus the potential energy will decrease. On the other hand, if this energy stays within the system, it should be transfered to some constituent modes within the star. Note that with the increase in $ N $, the number of neutrons, the kinetic energy increases as, $ N^{4/3}$, but the potential energy decreases as $ N^2$. There are hardly any constituent in significant numbers to carry the excess energy. This is where the gravitons become important. We have already seen that the soft gravitons collectively posses energy comparable to the potential energy of the star.\\
In refeference \cite{dyr,xulu}, this problem has been studied for critical neutron stars within the framework of general relativity. 
Using various definitions of  gravitational energy  within general relativity, the authors find  that  30  to 40 percent of energy content of star is in the form of gravitational field. This is consistent with our heuristic argument. Along with our arguments for the dominance of system size interaction, this again suggest that for understanding compact stars such as neutron stars and black holes, we should be studying the role played by gravitons of system size wavelength. Like the gluons in QCD (and unlike the photons in QED),
the gravitons have self-interaction and are likely to  form collective state .\\
The Dvali-Gomez conjecture states that the number of such system size gravitons is maximum for black hole, and they are self sourcing.
% in the sense that 
The kinetic energy and the potential energy of self-interaction approximately balance each other. They also suggest that, at maximum packing within the black hole, the system size gravitons undergo transition to bose condensed state \cite{dval2}. In the light of the discussion above, such a scenario for the black holes clearly look very promising. \\
Thus even for black holes, it is the long range part of the gravity thet plays the major role. This leads us to conclude that the essence of gravity 
lies in the long range nature of the interaction.

%Your text goes here…

% Uncomment the following two lines if you want to have a bibliography
%\bibliographystyle{alpha}
%\bibliography{document}

\begin {thebibliography}{99}
\bibitem{thir}    W. Thirring,  A Course of Mathematical Physics, Vol 4, Springer, 1983
\bibitem{lieb1}  E.H. Lieb and H.-T. Yau,   Phys. Rev. Lett. {\bf 61} (1988) 1695; Commun. Math. Phys. {\bf 118} (1988) 177
\bibitem{lieb2}  E.H. Lieb, M. Loss and H. Siedentop, Helv. Phys. Acta {\bf 69} (1996) 974 ; arXiv: 9608060[{\bf v1} cond-mat]
\bibitem{cohe}  B.L. Cohen, Concepts of Nuclear Physics, McGraw Hill, 1971
\bibitem{shal}   A. deShalit and H. Feshbach, Theoretical Nuclear Physics, Vol. 1, Chapters I and II, John Wiley and Sons, Inc., 1974
\bibitem{jean}   James H. Jeans, Phil. Trans. Roy. Soc. (London) {\bf 199} (1902) 1
\bibitem{mich}  Michael K.-H. Kiessling, Adv. Appl. Math. {\bf 31} (2003) 132
\bibitem{clar}     C. Clarke and B. Carlswell, Astrophysical Fluid Dynamics, Cambridge University Press, Cambridge, 2007
\bibitem{shap}   S.L. Shapiro and S.A. Teukolsky,    Black Holes, White Dwarfs and Neutron Stars: The Physics of Compact Objects, 
                                John Wiley, New York, 1983
\bibitem{chan1} S. Chandrasekhar,  Nobel Lecture, The Nobel Foundation, 1984, Published in  Rev. Mod. Phys. {\bf 56} (1984) 137
\bibitem{adel}  E.G. Adelberger, B.R. Heckel and A.E. Nelson, Ann. Rev. Nucl. Part. Sci. {\bf 53} (2003) 77, arXiv: 0307284{\bf v1}  [gr-qc]
\bibitem{salu}  E.J. Salumbides, W. Ubachs and V.I. Korobov, Journal of Molecular Spectroscoph, {\bf 300} (2014) 55, arXiv: 1308.1711{\bf v1}  [hep-ph]
\bibitem{azam} M. Azam, M. Sami, C.S. Unnikrishnan and T. Shiromizu, Phys. Rev. ( Rapid Communication) {\bf D77} (2008) 101101
\bibitem{azam1} M.Azam and M. Sami, Phys. Rev. {\bf D72} (2005) 024024
%Hydrodynamic and Hydrodynamic Stability, Oxford University Press, Oxford,  1961
%\bibitem{lieb3}   E.H. Lieb and H.-T. Yau,  Commun. Math. Phys. {\bf 112} (1987) 147
\bibitem{dval1}  G. Dvali and C. Gomez, "Self-completeness of Einstein Gravity",  arxiv:1005.3497{\bf v1} [hep-th]
\bibitem{dval2}   G. Dvali and C. Gomez, "Black Holes Quantum N-Portrait",  Fortsch. Phys. {\bf 61} (2013) 742 ; arxiv: 1112.3359{\bf v1} [hep-th]
\bibitem{dyr}   D. Dyrda, B. Kinasiewisz, M. Kutschera and M. Szmaglinski, Acta Phys. Polon.{\bf B 37} (2006) 357, arXiv: 0601061{\bf v1} [gr-qc]
\bibitem{xulu}  S.S. Xulu, PhD thesis, arXiv: 0308070{\bf v1}  [hep-th]






\end{thebibliography}

\end{document}